\documentclass[12pt]{article}

\def\LA{\langle} % Left angle
\def\RA{\rangle} % Right angle
\def\B<#1|{\mathopen{\LA}{#1}\mathclose{|}} % bra
\def\K|#1>{\mathopen{|}{#1}\mathclose{\RA}} % ket
\def\BK<#1|#2>{\mathopen{\LA}{#1}\mathbin{|}{#2}\mathclose{\RA}} % <|>
\def\ME<#1|#2|#3>%
  {\mathopen{\LA}{#1}\mathbin{|}{#2}\mathbin{|}{#3}\mathclose{\RA}} % <| |>

\def\vec#1{\bf #1}

\title{Norm Kernels and the Closeness Relation for Pauli-Allowed Basis Functions}

\author{G.~F.~Filippov$^{1)}$, Yu.~A.~Lashko$^{1,2)}$, S.~V.~Korennov$^{1,2)}$, and
K.~Kat\=o$^{2)}$\\
{\normalsize $^{1)}$Bogolyubov Institute for Theoretical Physics,
14-b
Metrolohichna Str.}\\
{\normalsize Kiev-143, Ukraine }\\
{\normalsize $^{2)}$Graduate School of Science, Hokkaido
University, Sapporo 060-0810, Japan}}

\begin{document}

\maketitle

\begin{abstract}
The norm kernel of the generator-coordinate method is shown to be
a symmetric kernel of an integral equation with eigenfunctions
defined in the Fock--Bargmann space and forming a complete set of
orthonormalized states (classified with the use of SU(3) symmetry
indices) satisfying the Pauli exclusion principle. This
interpretation allows to develop a method which, even in the
presence of the SU(3) degeneracy, provides for a consistent way to
introduce additional quantum numbers for the classification of the
basis states. In order to set the asymptotic boundary conditions
for the expansion coefficients of a wave function in the SU(3)
basis, a complementary basis of functions with partial angular
momenta as good quantum numbers is needed. Norm kernels of the
binary systems $^6$He$+p$, $^6$He$+n$, $^6$He$+^4$He, and
$^8$He$+^4$He are considered in detail.
\end{abstract}

\section{Introduction}

The concept of norm kernel in the generator-coordinate space was
introduced by Horiuchi\cite{Hor70} and has been important in the
realization of the resonating-group method (RGM)\cite{MRG}. In
order to find it, two Slater determinants are constructed on the
Bloch--Brink orbitals\cite{BB} depending on two (different, in
general) sets of vector generator parameters as well as the
single-particle coordinates. Each of them generates a basis of the
harmonic-oscillator states allowed by the Pauli principle. The
norm kernel is obtained by integrating the product of the
determinants over all single-particle variables. It is thus a
function of the generator parameters which reproduces the relative
motion of the nuclear clusters with the internal wave functions of
the clusters being fixed.

A Pauli-allowed basis state defined in the coordinate space can be
mapped onto the generator-parameter space by expansion of the norm
kernel over the powers of the generator parameters. As the number
of the generator parameters is relatively small, the map appears
to be much simpler than the original. Also, the norm kernel
contains the eigenvalues of the basis functions.

The norm kernels of a number of nuclear systems were studied in
works of Horiuchi and Fujiwara\cite{HorFud}. They have shown that
the diagonalization of the norm kernel requires the basis to be
labeled with the quantum indices $(\lambda,\mu)$ of irreducible
representations  of the SU(3) group ({\it SU(3) irreps}). The
eigenvalues of the norm kernel depend on the total number of the
oscillator quanta and $(\lambda,\mu)$, and do not depend on the
angular momenta of the basis states. In ref.\cite{HorFud}, the
diagonal form of the norm kernel for a two-cluster system, if at
least one of the clusters has valence neutrons, depends on the
SU(3) Clebsch--Gordan coefficients. They appear if the reduction
SU(3)$\otimes$SU(3)$\supset$SU(3) is needed.

Usually the generator parameters are taken to be real which makes
their physical interpretation easy. However, in this case the
generator-coordinate method (GCM) meets some difficulties due to
the singularity of the Gaussian transformation used in the
Griffin--Hill--Wheeler theory\cite{BrW}. This difficulty is
avoided when the generator parameters are considered to be
complex\cite{BrW},\cite{Wong}. Besides, there are other merits of
the analytic continuation of the generator parameters onto the
complex space. Below we show that in this case the norm kernel can
be considered as a symmetric kernel of the Fredholm integral
equation of the second kind. Then, its eigenfunctions (which are
possible to find analytically) become the Pauli-allowed
harmonic-oscillator basis states in the Fock--Bargmann
representation, normalized with the Bargmann measure, while its
eigenvalues appear to be directly related to those of the norm
kernel.

The Hilbert--Schmidt expansion of the norm kernel is naturally
interpreted as the closeness condition for the set of
Pauli-allowed basis states in the RGM. These states are
straightforwardly obtained in the Fock--Bargmann representation as
long as there is no SU(3) degeneracy. If there is one, these
functions are found as solutions of an integral equation with the
degenerate kernel.

The method of projection of a direct product of SU(3) irreps onto
the states having a definite SU(3) symmetry described here does
not require the explicit use of the SU(3) Clebsch--Gordan
coefficients and directly yields the SU(3) weight factors. In the
presence of the degeneracy, the method provides for a consistent
way to introduce additional quantum numbers for the classification
of the basis states.

A drawback of the basis of SU(3) irreps ({\it SU(3) basis}) is
that, when a wave function of continuous spectra is expanded over
its states, the expansion coefficients have a complicated form in
the asymptotic limit of large number of oscillator quanta. In
order to find the asymptotic behavior of these coefficients, a
relation should be established between the SU(3) basis and the
"{\it physical}" basis, the latter having the partial angular
momenta of clusters and of their relative motion as its quantum
numbers. The SU(3) basis functions are expanded over the
"physical" basis states with the use of the integration technique
in the Fock--Bargmann space developed by the present authors.

The issues covered in this paper are closely related to the
problem of scattering of neutron-rich nuclei studied in recent
experiments with radioactive nuclear beams. In cases when nuclei
with open shells are involved, the RGM norm kernel becomes rather
complicated and its analysis requires new approaches.

We begin with explaining our method on a well-known example of two
interacting alpha-particles. Then we proceed to other cases of
systems composed of a $p$-cluster and an $s$-cluster: $^6$He$+p$,
$^6$He$+n$, $^6$He$+^4$He, and $^8$He$+^4$He.

\section{Norm Kernel for $^8$Be}

Important advantages of the Fock--Bargmann representation become
clear even in the simple case of $^8$Be=$^4$He+$^4$He. Various
methods have been tested on this system. First, we briefly review
an algorithm of the construction of the two-cluster norm kernel of
$^8$Be described by Saito \cite{Saito77}, and then we compare it
with another approach, in which the Fock--Bargmann representation
is consistently used.

In \cite{Saito77}, the Slater determinant $\bar{\Phi}$ was
constructed from four single-particle states $\psi_1({\vec r}_i)$
in the harmonic-oscillator field with the origin at ${\vec R}_1$
\begin{equation}
\label{a1} \psi_1({\vec r}_i)={1\over\pi^{3/4}}
\exp\left(-{1\over2}({\vec r}_i-{\vec R}_1)^2\right),~~i=1,2,3,4,
\end{equation}
and four more states $\psi_2({\vec r}_i)$ in the field with the
center at ${\vec R}_2$
\begin{equation}
\label{a2} \psi_2({\vec r}_i)={1\over\pi^{3/4}}
\exp\left(-{1\over2}({\vec r}_i-{\vec R}_2)^2\right),~~i=5,6,7,8.
\end{equation}
These single-particle states are the so-called {\it Bloch--Brink
orbitals} which are often used in the alpha-particle model of the
nucleus, that is, when a multi-center approximation is natural.
The wave equation is then written not in the coordinate space, but
in a representation where the wave function depends on the
generator coordinates\cite{GWh} ${\vec R}_i$ -- the vectors
describing the motion of the centers of the clusters. As the
internal cluster functions are fixed and chosen to have a simple
form, the number of independent variables is reduced and the
attention is drawn to dynamics of the relative motion of the
clusters.

A similar determinant $\tilde{\bar{\Phi}}$ is constructed from the
{\it conjugated} orbitals
\begin{equation}
\label{a4a}
 \tilde{\psi}_1({\vec
r}_i)={1\over\pi^{3/4}} \exp\left(-{1\over2}({\vec r}_i-{\vec
S}_1)^2\right),~~i=1,2,3,4,
\end{equation}
\begin{equation}
\label{a4b} \tilde{\psi}_2({\vec r}_i)={1\over\pi^{3/4}}
\exp\left(-{1\over2}({\vec r}_i-{\vec S}_2)^2\right),~~i=5,6,7,8.
\end{equation}

The nature of the resonating-group method is variational.
Therefore the construction of the overlap integral
$\BK<\tilde{\bar{\Phi}}|\bar{\Phi}>$ and the energy functional
$\ME<\tilde{\bar{\Phi}}|\hat{H}|\bar{\Phi}>$ for the Hamiltonian
$\hat{H}$ of the nucleus $^8$Be is followed by the independent
variation of the orbitals $\psi_1({\vec r}_i),~\psi_2({\vec r}_i)$
and $\tilde{\psi}_1({\vec r}_i),~\tilde{\psi}_2({\vec r}_i)$.

It is convenient to modify the Bloch--Brink orbitals
(\ref{a1})--(\ref{a2}) as follows,
\begin{equation}
\label{a6a} \phi_1({\vec r}_i)={1\over\pi^{3/4}}
\exp\left(-{1\over2}{\vec r}^2_i+\sqrt{2}({\vec R}_1 \cdot{\vec
r}_i)- {1\over2}{\vec R}_1^2\right),~~i=1,2,3,4,
\end{equation}
\begin{equation}
\label{a6b} \phi_2({\vec r}_i)={1\over\pi^{3/4}}
\exp\left(-{1\over2}{\vec r}^2_i+\sqrt{2}({\vec R}_2\cdot{\vec
r}_i)- {1\over2}{\vec R}_2^2\right),~~i=5,6,7,8.
\end{equation}
The {\it modified Brink orbital} is the kernel of
the transformation from the coordinate space to the space of
complex generator parameters (the Fock--Bargmann
space\cite{Barg}). It has other useful properties, in particlular,
it is the generating function for the harmonic-oscillator
basis\cite{HorGen},\cite{HorG}.

Another simplification is the replacement of the single-particle
variables ${\vec r}_i$ by the Jacobi vectors. As a result, in the
Slater determinants $\Phi$ and $\tilde{\Phi}$ constructed from the
modified Brink orbitals the wave functions of the center-of-mass
motion are factored out (and dropped out from now
on)\footnote{Each of the two determinants is a kernel of the
integral transform from the coordinate space to the Fock--Bargmann
space. In the RGM, this transform is carried with a reduction in
the number of independent variables.}.

The overlap integral $\BK<\tilde{\bar{\Phi}}|\bar{\Phi}>$
(integration is over all single-particle variables) is obtained in
the form
\begin{eqnarray}
\label{b1} I_S({\vec R},{\vec
S})=\exp\left(-{\vec{R}^2\over2}\right)8 \sinh^4{\left({\vec R}
\cdot {\vec S}\right)\over4} \exp\left(-{\vec{S}^2\over2}\right),
\end{eqnarray}
where
$${\vec R}_1=-{\vec R}_2={{\vec R}\over2},~~{\vec S}_1=-{\vec S}_2={{\vec S}\over2}.$$
In the orbitals are modified, the result is somewhat simpler,
\begin{eqnarray}
\label{b2} I({\vec R},{\vec S})=8\sinh^4{\left({\vec R} \cdot
{\vec S}\right)\over4}.
\end{eqnarray}
Eqs. (\ref{b1}) and (\ref{b2}) are symmetric norm kernels.

\section{Properties of Norm Kernels}

A norm kernel can, in principle, be expanded over a complete set
of its eigenfunctions. The expanded form of (\ref{b1}) and
(\ref{b2}), respectively, is
\begin{eqnarray}
\label{d1} I_S({\vec R},{\vec S})=
\sum_n\bar{\lambda}_n\bar{\psi}_n({\vec R})\bar{\psi}_n({\vec S}),
\end{eqnarray}
\begin{eqnarray}
\label{d2} I({\vec R},{\vec S})= \sum_n\lambda_n\psi_n({\vec
R})\psi_n({\vec S}),
\end{eqnarray}
where $\bar{\lambda}_n$ and $\lambda_n$ are the eigenvalues,
$\bar{\psi}_n({\vec R})$ and $\psi_n({\vec R})$ are the
eigenvectors. These equations serve as closeness conditions for
the bases $\{\bar{\psi}_n({\vec R})\}$ and $\{\psi_n({\vec R})\}$
of allowed states in a space of the generator vector ${\vec R}$.

The norm kernels (\ref{b1}) and (\ref{b2}) do differ. To
comprehend the nature of this difference, we need to match them
with integral equations having the orthonormalized states
$\bar{\psi}_n({\vec R})$ and $\psi_n({\vec R})$, respectively, as
their solutions. But first, we need to define the range of
integration in the generator parameter space as well as the
measure.

In Saito's treatment, the generator parameters ${\vec R}_k$ have a
simple physical interpretation. They describe spatial translations
of the cluster centers, and thus their Cartesian components
$R_{kx},R_{ky},R_{kz}$ are in the range
$$-\infty < R_{kx},R_{ky},R_{kz} < \infty.$$
If the measure is set to the unity, one arrives to the integral
equation
\begin{eqnarray}
\label{c1} \lambda\bar{\psi}({\vec
R})=\int\exp\left(-{\vec{R}^2\over2}\right) \sinh^4\left({{\vec R}
\cdot {\vec S}\over4}
\right)\exp\left(-{\vec{S}^2\over2}\right)\bar{\psi}({\vec
S})d{\vec S}.
\end{eqnarray}

Consider now the case of the modified orbitals,
Eqs.(\ref{a6a})--(\ref{a6b}). Each of them is an eigenfunction of
the position vector operator $\hat{\vec r}_i$
\begin{eqnarray}
\hat{\vec r}_i={1\over\sqrt{2}}\left({\vec R}_k+\vec\nabla_{{\vec
R}_k}\right),~~ k=1,2,
\end{eqnarray}
defined in the Fock--Bargmann space with the eigenvalue ${\vec
r}_i$. The independent variables in the Fock--Bargmann space are
complex vectors
\begin{eqnarray}
{\vec R}_k={\overrightarrow{\xi}_k+i\overrightarrow{
\eta}_k\over\sqrt{2}},~~k=1,2.
\end{eqnarray}
Here $\overrightarrow{\xi}_k$ and $\overrightarrow{\eta}_k$ are
coordinate and momentum vectors, respectively. In fact, functions
of ${\vec R}_k$ are defined in the phase space. In general, the
number of vectors ${\vec R}_k$ depends on the assumed cluster
structure of the system under consideration.

The integral equation in the Fock--Bargmann space takes the form
\begin{eqnarray}
\label{c2} \lambda\psi({\vec R})=\int8\sinh^4{\left({\vec R}\cdot
\bar{\vec S}\right)\over4} \psi(\bar{\vec R}) d\mu_b,
\end{eqnarray}
where
$$d\mu_b=\exp\{-(\bar{\vec R} \cdot \bar{\vec
S})\}{d\vec{\bar{\xi}}d\vec{\bar{\eta}}\over(2\pi)^3}$$ is the
{\it Bargmann measure}.

The solutions of Eqs.(\ref{c1}),(\ref{c2}) should be presented as
Hilbert--Schmidt expansions (\ref{d1}--\ref{d2}) of the respective
kernels. The latter expansion, Eq.(\ref{d2}), is of particular
interest. It is a {\it density matrix} defined in the phase space,
and, as such, can be directly used in quantum statistics.

As for Eq.(\ref{c1}), due to the well-known problem of singularity
of the weight functions in the generator-coordinate method with
real parameters, it cannot be easily solved. The eigenvalues of
the kernel were found in ref. \cite{Saito77} by expanding it in a
series over the powers of the generator parameters. However, the
eigenfunctions of (\ref{b1}) do not enter this expansion and thus,
it is not the Hilbert--Schmidt expansion (\ref{d1}) for this
kernel. On the other hand, the solution of Eq.(\ref{c2}) (both
eigenvalues and eigenfunctions of the kernel (\ref{b2})) is
readily found. It suffices to expand the kernel (\ref{b2}) in a
series over powers of the dot product $({\vec R} \cdot {\vec S})$
and make sure that a $2n$-th term (only even powers appear in the
expansion) is a packet of states belonging to the $(2n,0)$ irrep.
Indeed, in the expansion
\begin{eqnarray}
I({\vec R},{\vec S})= \sum_{n=2}^\infty
{1\over(2n)!}\left(1-{4\over2^{2n}}\right) ({\vec R} \cdot {\vec
S})^{2n}
\end{eqnarray}
the expression
\begin{eqnarray}
\label{c3} I_{(2n,0)}={1\over(2n)!}({\vec R} \cdot {\vec S})^{2n}
\end{eqnarray}
is a norm kernel for the irrep $(2n,0)$. It is important that
\begin{eqnarray}
\label{e1} \int {1\over(2n)!}({\vec R} \cdot {\vec S})^{2n}
d\mu_b={(2n+1)(2n+2)\over2},
\end{eqnarray}
that is, integration of the kernel $I_{(2n,0)}$ with the Bargmann
measure\footnote{See Appendix A.} yields the number of the basis
states belonging to the SU(3) irrep $(2n,0)$ in accordance with
the formula\cite{Harvey}
\begin{equation}
\label{dimensionality}
 \dim [\lambda\mu] = {(\lambda+1)(\mu+1)(\lambda+\mu+2)
\over 2}.
\end{equation}
The choice of a common norm factor at the kernel $I({\vec R},{\vec
S})$ worths some attention. In the example considered here, $n$ is
the total number of oscillator quanta. The normalization of the
kernel to the {\it dimensionality of the SU(3) irrep} adopted here
is consistent with the statistical interpretation of $I({\vec
R},\vec S)$ as a density matrix.

The wave packet (\ref{c3}) can be written down as a sum of
overlaps of states with definite values of the number of quanta
$2n$, angular momentum $l$ and its projection $m$
\begin{eqnarray}
\label{c4} I_{(2n,0)}=\sum_{l,\,m}\psi_{2n,\,l,\,m}({\vec
R})\psi_{2n,\,l,\,m}({\vec S}),
\end{eqnarray}
where $l$ takes all possible even values from 0 to $2n$, and basis
function $\psi_{2n,\,l,\,m}({\vec R})$ are normalized to unity,
\begin{eqnarray}
\label{c5} \int\psi_{2n,\,l,\,m}({\vec R})\psi_{2n,\,l,\,m}({\vec
S})d\mu_b=1.
\end{eqnarray}

Some of the simplest basis functions (normalized using the
formulae in Appendix A) are:
\begin{eqnarray*}
\psi_{0,\,0,\,0}=1,~\psi_{2,\,0,\,0}={1\over\sqrt{3!}}{\vec R}^2,~
\psi_{2n,\,0,\,0}={1\over\sqrt{(2n+1)!}}{\vec R}^{2n},
\end{eqnarray*}
\begin{eqnarray}
\label{ps2n0} \psi_{2,\,2,\,0}= \sqrt{{2\over3!}}\left({3\over2}{
R}^2_z-{1\over2}{\vec R}^2\right).
\end{eqnarray}

Now the expansion of the norm kernel (\ref{b2}) takes the form
\begin{eqnarray}
I({\vec R,\vec S})= \sum_{n=2}^\infty
\left(1-{4\over2^{2n}}\right) \sum_{l,\, m}\psi_{2n,\, l,\,
m}({\vec R})\psi_{2n,\, l,\, m}({\vec S}).
\end{eqnarray}
Thus we have found the complete set of the orthonormalized
eigenfunctions $\{\psi_{2n,\,l,\,m}({\vec R})\}$ and eigenvalues
$$\lambda_{2n,\,l,\,m}=1-{4\over2^{2n}}$$ of the norm kernel
(\ref{b2}) without solving the integral equation (\ref{c2})
directly.

Note that the eigenvalues of (\ref{b2}) limit to a finite value
(unity) at $n\rightarrow\infty$, thus the kernel itself is of
special kind\cite{Petr}. The deviations from the unity are due to
the Pauli exclusion principle.

\section{Generalizations}

The example of the $\alpha+\alpha$ system considered above is of
heuristic nature and can lay ground for some generalizations to
more complex systems studied within the framework of the RGM.

\subsection{Closeness Condition}

It is known that the closeness condition for a set of basis
function in the coordinate space, $\{\varphi_n(\vec{r})\}$, is
\begin{eqnarray}
\label{u1} \delta({\vec r}-{\vec
r}')=\sum_n\varphi_n(\vec{r})\varphi^*_n(\vec{r}'),
\end{eqnarray}
where the sum runs over all values of the quantum numbers $n$. In
the Fock--Bargmann space, this condition reads
\begin{eqnarray}
\label{u2} \exp({\vec R \cdot \vec S})=\sum_n\psi_n({\vec
R})\psi_n({\vec S}),
\end{eqnarray}
where $\psi_n({\vec R})$ is the Fock--Bargmann map of
$\varphi_n(\vec{r})$. A unitary transform of the set
$\varphi_n(\vec{r})$ or its map $\psi_n({\vec R})$ leaves Eqs.
(\ref{u1}--\ref{u2}) intact. On the other hand, solution of the
integral equation
\begin{eqnarray}
\label{u3} \lambda\Psi({\vec R})=\int I({\vec R},{\bar{\vec
S}})\Psi({\bar{\vec R}})d\mu_b
\end{eqnarray}
yields an orthonormalized set $\{\Psi_n({\vec R})\}$ as well, but
these states are allowed by the Pauli principle, and the closeness
relation for them is expressed by the Hilbert--Schmidt expansion,
\begin{eqnarray}
\label{u4} \BK<\tilde{\Phi}|\Phi> \equiv I({\vec R},{\vec
S})=\sum_n\lambda_n\Psi_n({\vec R})\Psi_n({\vec S}),
\end{eqnarray}
which is quite different from the expansion (\ref{u2}). Indeed,
whereas Eq.(\ref{u2}) is valid for various sets of orthonormalized
functions, as long as those sets are complete, only uniquely
defined eigenfunctions of the symmetric kernel $I({\vec R},{\vec
S})$ enter Eq.(\ref{u4}).

Although the norm kernel here, in fact, was expanded over a
complete set of the eigenstates $\Psi_n({\vec R})$ of the
antisymmetrization operator, it may be alternatively expanded over
eigenfunctions of a Hamiltonian. But such an expansion should be
performed carefully. Prior to it, the functions $\Psi_n({\vec R})$
have to be replaced with ${\bar{\Psi}}_n({\vec R})$ defined by
\begin{eqnarray}
\label{u5} {\bar{\Psi}}_n({\vec R})=\sqrt{\lambda_n}\Psi_n({\vec
R}).
\end{eqnarray}
The new basis functions are still orthogonal but have an
additional norm factor $\sqrt{\lambda_n}$.
Eqs.(\ref{u4}--\ref{u5}) are followed by
\begin{eqnarray}
\label{u6} I({\vec R},{\vec S})= \sum_n{\bar{\Psi}}_n({\vec
R}){\bar{\Psi}}_n({\vec S}).
\end{eqnarray}
The quadratic form (\ref{u6}) remains diagonal whatever unitary
transform is applied to the basis $\{{\bar{\Psi}}_n({\vec R})\}$.
One particular example of such unitary transform appears in the
algebraic version of the RGM (AV RGM)\cite{AVMRG}, where
eigenfunctions $F_\nu({\vec R})$ of the RGM Hamiltonian, defined
in the Fock--Bargmann space, are constructed in the form
\begin{eqnarray}
\label{uu8} F_\nu({\vec R})=\sum_{n}C^\nu_n{\bar{\Psi}}_n({\vec
R}),
\end{eqnarray}
both in discrete and continuous sections of the spectrum. The set
of coefficients $C^\nu_n$ is a unitary matrix, thus the expansion
(\ref{u6}) is replaced with another one, containing an integration
over continuum states $F_E({\vec R})$ with energy $E$ and a sum
over discrete states $F_\nu({\vec R})$.
\begin{eqnarray}
\label{u9} I({\vec R},{\vec S})= \sum_{\nu=1}^{\nu_0} F_\nu({\vec
R})F_{\nu}({\vec S})+ \int_0^\infty F_E({\vec R})F_E({\vec S})dE.
\end{eqnarray}
The coefficients $C_n^\nu$ and $C_n^E$ are computed in the
following way. The overlap of the Slater determinants with the RGM
Hamiltonian is expanded over either the states $\Psi_n({\vec R})$,
\begin{eqnarray}
\label{v7} \ME<\tilde{\Phi}|\hat{H}|\Phi>=\sum_{n,\tilde{n}}
\tilde{\Psi}_{\tilde{n}}({\vec
S})\ME<\tilde{n}|\hat{H}|n>\Psi_n({\vec R}),
\end{eqnarray}
($\ME<\tilde{n}|\hat{H}|n>$ are the matrix elements of the
Hamiltonian), or the states ${\bar{\Psi}}_n({\vec R})$,
\begin{eqnarray}
\label{v8} \ME<\tilde{\Phi}|\hat{H}|\Phi>=\sum_{n,\tilde{n}}
{\bar{\Psi}}_{\tilde{n}}({\vec S})
{\ME<\tilde{n}|\hat{H}|n>\over\sqrt{\lambda_n\lambda_{\tilde{n}}}}{\bar{\Psi}}_n({\vec
R}).
\end{eqnarray}
After that, the following quadratic form is diagonalized,
\begin{eqnarray}
\label{v9}
\ME<\tilde{\Phi}|\hat{H}-E|\Phi>=\sum_{n,\tilde{n}}{\bar{\Psi}}_{\tilde{n}}({\vec
S})
\left({\ME<\tilde{n}|\hat{H}|n>\over\sqrt{\lambda_n\lambda_{\tilde{n}}}}
-E\delta_{n,\tilde{n}}\right){\bar{\Psi}}_n({\vec R}).
\end{eqnarray}
This is done by a transition from the basis states
${\bar{\Psi}}_n({\vec R})$ to $F_\nu({\vec R})$ and $F_E({\vec
R})$, which transforms Eq.(\ref{v9}) into
\begin{eqnarray}
\label{v10}
\ME<\tilde{\Phi}|\hat{H}-E|\Phi>&=&\sum_{\nu=1}^{\nu_0}
F_\nu({\vec R})\left(E_\nu-E\right)F_{\nu}({\vec S}) \nonumber \\
&+&\int_0^\infty F_{E'}({\vec R})\left(E'-E\right)F_{E'}({\vec
S})dE'.
\end{eqnarray}
This brings us to the spectrum of the RGM Hamiltonian in the
Fock--Bargmann space, and in the representation of the
Pauli-allowed harmonic-oscillator states.

The norm kernel can be treated as an RGM density
matrix\cite{Landau} in the phase space. From its structure we
conclude that it is the density matrix of a "mixed"
state\footnote{The spectrum of eigenvectors and eigenvalues of a
norm kernel (a density matrix) is uniquely defined. This also
unambiguously defines those operators which the Hamiltonian
generating this spectrum is composed of. However, the norm kernel
{\it itself} is not unique; its form depends on the choice of
generator parameters.}. It takes a diagonal form only in the
representation of the basis of its eigenfunctions. A unitary
transform of the basis disrupts the diagonal form of the "mixed"
density matrix. This would not be the case if all the eigenvalues
were equal to unity. Then the summation of the density matrix over
all allowed states would yield a Slater determinant of overlaps,
i.e.
$$\det||\exp({\vec R}_i \cdot {\vec S}_j)||,$$
where ${\vec R}_i$ is the complex vector parameter for $i$th
particle. Finally, the closeness relation for the set
$\{\Psi_n(\{{\vec R}_i\})\}$ in the Fock--Bargmann representation
would read\cite{Per}
\begin{eqnarray}
\sum_n\Psi_n(\{{\vec R}_i\})\Psi_n(\{{\vec S}_i\})=
\det||\exp({\vec R}_i \cdot {\vec S}_j)||.
\end{eqnarray}

\subsection{Eigenvalues of the Norm Kernel}

The Fock--Bargmann representation open a simple way to explain the
physical sense of the eigenvalues of the norm kernel. Its
eigenfunctions possess a permutational symmetry, since they are,
in fact, eigenfunctions of the antisymmetrization operator. Again,
consider the $\alpha+\alpha$ system as an example. The norm kernel
is
\begin{eqnarray}
I({\vec R},{\vec S})=\cosh({\vec R \cdot \vec S})-4\cosh\{({\vec R
\cdot \vec S})/2\}+3.
\end{eqnarray}
The antisymmetrization operator $\hat{A}$ is defined by the
following relation,
\begin{eqnarray}
I({\vec R},{\vec S})=\hat{A}~\cosh({\vec R \cdot \vec S}).
\end{eqnarray}
If $F({\vec R}^2)$ is an entire function of even powers of ${\vec
R}$, then
\begin{eqnarray}
\hat{A}~F({\vec R}^{2})=F({\vec R}^{2})-4F\left({{\vec
R}^2\over4}\right)+ F(0).
\end{eqnarray}
In particular,
\begin{eqnarray}
\hat{A}~{\vec R}^{2n}=\left(1-{1\over4^{n-1}}\right){\vec R}^{2n}.
\end{eqnarray}
Therefore, the eigenvalues of $\hat{A}$ are, expectedly, equal to
$$1-{1\over4^{n-1}},$$
if $n>1,$ and zero, if $n=0,~1$.

\section{Norm Kernels in the SU(3) Model}

Before proceeding to the RGM norm kernels in the case of open
$p$-shell clusters, we consider Elliott's SU(3) model\cite{Ell}
where it is also appropriate to define the norm kernels in the
Fock--Bargmann space. Their form is particularly simple for the
{\it leading} irreducible representations. In general, they depend
on two complex vectors ${\vec u}$, ${\vec v}$ and their complex
conjugate counterparts ${\vec u}^*$, ${\vec v}^*$. These vectors
are independent variables of wave functions of the valence
nucleons in the Fock--Bargmann representation. If the leading
SU(3) irrep is labelled by the indexes $(\lambda,\mu)$, then its
norm kernel $I_{(\lambda,\mu)}({\vec u},{\vec v};{\vec u}^*,{\vec
v}^*)$ is
\begin{eqnarray}
\label{q1} I_{(\lambda,\, \mu)}({\vec u},{\vec v};{\vec u}^*,{\vec
v}^*)= {\lambda+1\over (\lambda+\mu+1)!\mu!}({\vec u}\cdot {\vec
u}^*)^\lambda ([{\vec u}\times{\vec v}][{\vec u}^*\times{\vec
v}^*])^\mu.
\end{eqnarray}
The kernel (\ref{q1}) is normalized to the dimensionality
(\ref{dimensionality}) of the SU(3) irrep, i.e.
\begin{eqnarray}
\label{q2} \int I_{(\lambda,\, \mu)}({\vec u},{\vec v};{\vec
u}^*,{\vec v}^*)d\mu_b= {(\lambda+1)(\mu+1)(\lambda+\mu+2)\over2}.
\end{eqnarray}

The explicit form of the wave functions $\Psi_{L\alpha M}({\vec
u},{\vec v})$ in the Elliott's model can be derived from the
Hilbert--Schmidt expansion of the kernel (\ref{q1}),
\begin{eqnarray}
I_{(\lambda,\,\mu)}({\vec u},{\vec v};{\vec u}^*,{\vec v}^*)=
\sum_{L\alpha M}\Psi_{L\alpha M}({\vec u},{\vec v}) \Psi_{L\alpha
M}({\vec u}^*,{\vec v}^*)
\end{eqnarray}
The expansion itself is projection of the kernel to the states
with definite values of the angular momentum $L$, its projection
$M$ and, if necessary, an additional quantum number $\alpha$. The
latter is needed to label the states with the same $L$ in a given
SU(3) irrep, and only when such multiplicity exists does one need
to solve an integral equation with the degenerate kernel. As an
illustration, we show a normalized wave function for $L=0$ (both
$\lambda$ and $\mu$ are even),
\begin{eqnarray}
\Psi_{0}({\vec u},{\vec v})=N_0 \left({\vec u \cdot \vec
u}\right)^{\lambda/2} \left([{\vec u}\times{\vec
v}]^2\right)^{\mu/2},
\end{eqnarray}
\begin{eqnarray*}
N_0^2={1\over2\sqrt{\pi}}\cdot{\lambda+1\over
(\lambda+\mu+1)!\mu!}
{\Gamma((\lambda+1)/2)\Gamma((\lambda+\mu)/2+1)\Gamma((\mu+1)/2)\over
\Gamma(\lambda/2+1)\Gamma((\lambda+\mu+3)/2)\Gamma(\mu/2+1)}.
\end{eqnarray*}

\section{Norm Kernels for Some Two-Cluster Systems}

If a scattering of two open-shell clusters in the leading SU(3)
representation ($(\lambda_1,\mu_1)$ for the first cluster,
$(\lambda_2,\mu_2)$ for the second) is studied in the
Fock--Bargmann space, the norm kernel depends on five complex
vectors. Two of them, ${\vec u}_1$ and ${\vec v}_1$, describe the
degrees of freedom of the first cluster, two more, ${\vec u}_2$
and ${\vec v}_2$, are needed for the second one, and the fifth
vector ${\vec R}$ describes the relative motion of the clusters.

The nucleons in the $s$-shell are described by the modified Brink
orbital (\ref{a6a}), while for those in the $p$-shell the form of
the orbital is
\begin{eqnarray}
\label{a6-p}
 \phi_i({\vec r})={1\over\pi^{3/4}}\sqrt{2}({\vec
u}_i\cdot {\vec r}) \exp\left(-{{\vec r}^2\over2}+\sqrt{2}({\vec R
\cdot \vec r})-{{\vec R}^2\over2}\right),
\end{eqnarray}

The norm kernel takes the form of a sum of terms
$$I_n({\vec u}_1,{\vec v}_1,{\vec u}_2,{\vec v}_2,{\vec R};
{\vec u}^*_1,{\vec v}^*_1,{\vec u}^*_2,{\vec v}^*_2,{\vec S})$$
with definite number $n$ of oscillator quanta along ${\vec R}$.

\subsection{System $^6$He$+p$}

For the system $^6$He$+p$, two vectors, ${\vec u}$ and ${\vec R}$,
are needed. The first vector is for the open-shell cluster $^6$He,
the second is for the relative motion of the clusters. The $^6$He
has two neutrons in the $p$-shell which are described by an
orbital (\ref{a6-p}) with the vector ${\vec u}$. The norm kernel
is written as
\begin{eqnarray}
\label{q0} I({\vec u},{\vec R};{\vec u}^*,{\vec S})=\sum_n
\lambda_n I_n({\vec u},{\vec R};{\vec u}^*,{\vec S}),
\end{eqnarray}
\begin{eqnarray*}
\lambda_n=1-\left(-{1\over6}\right)^n,~~I_n({\vec u},{\vec
R};{\vec u}^*,{\vec S})= {1\over2\cdot n!}({\vec u \cdot \vec
u}^*)^2({\vec R \cdot \vec S})^n.
\end{eqnarray*}

The expression $I_n({\vec u},{\vec R};{\vec u}^*,{\vec S})$ is a
product of norm overlaps of two SU(3) irreps, namely, $(2,0)$ and
$(n,0)$, which is reducible to SU(3) irreps $(n+2,0),~(n,1)$ and
$(n-2,2)$. Total number of states (dimensionality) of the direct
product exactly equals to the sum of dimensionalities of the
latter three irreps,
$${(n+3)(n+4)\over2}+(n+1)(n+3)+{3(n-1)(n+2)\over2} = 3(n+1)(n+2). $$

The next problem is the separation of $I_n({\vec u},{\vec R};{\vec
u}^*,{\vec S})$ into a sum of overlaps for a definite SU(3)
irreps. It can be solved by introduction of the SU(3) group
generators $C_{i,j}$ expressed in terms of Cartesian components of
the vectors ${\vec u}$ and ${\vec R}$,
\begin{eqnarray}
C_{i,j}=u_i{\partial\over\partial u_j}+R_i{\partial\over\partial
R_j} -{1\over3}\delta_{i,j}\{({\vec u}\cdot \vec \nabla_{\vec
u})+({\vec R}\cdot \vec \nabla_{\vec R})\},
\end{eqnarray}
and the second-order Casimir operator of the SU(3) group,
\begin{eqnarray}
\hat{G}_2=\sum_{i,j}C_{i,j}C_{j,i}&=& ({\vec u}\cdot
\vec\nabla_{\vec u})+({\vec R}\cdot \vec\nabla_{\vec R}) \nonumber \\
&+&{2\over3}\{({\vec u}\cdot \vec\nabla_{{\vec u}})({\vec u}\cdot
\vec\nabla_{{\vec u}})+ ({\vec R}\cdot \vec\nabla_{{\vec
R}})({\vec R}\cdot
\vec\nabla_{{\vec R}}) \nonumber \\
&+&({\vec u}\cdot \vec\nabla_{{\vec R}})({\vec R}\cdot
\vec\nabla_{{\vec u}})+ ({\vec R}\cdot \vec\nabla_{{\vec
u}})({\vec u}\cdot \vec\nabla_{{\vec R}})\}
\end{eqnarray}
Acting onto $I_n({\vec u},{\vec R};{\vec u}^*,{\vec S})$ by
$\hat{G}_2$ iteratively\footnote{In practice, it suffices to act
with the operator
\begin{eqnarray*}
\hat{G}'_2=({\vec u}\cdot \vec \nabla_{{\vec R}})({\vec R}\cdot
\vec  \nabla_{{\vec u}})
\end{eqnarray*}
with the same final result.}, one can obtain three its
eigenfunctions, which are the norm kernels of three irreps:
\begin{eqnarray}
\label{irrep1} \BK<(n+2,0)|(n+2,0)>&=&{1\over2\cdot
n!}\left\{({\vec u \cdot \vec u}^*)^2({\vec R \cdot \vec S})^n
\phantom{.\over .} \right. \nonumber \\&-&{2n\over n+2}({\vec u
\cdot \vec u}^*)([{\vec u \times \vec R}]\cdot[{\vec u}^*\times
{\vec S}])({\vec R \cdot \vec S})^{n-1}
 \\
&+& \left.{n(n-1)\over (n+1)(n+2)}([{\vec u \times \vec R}]\cdot
[{\vec u}^*\times {\vec S}])^2 ({\vec R \cdot \vec
S})^{n-2}\right\}\nonumber
\end{eqnarray}
\begin{eqnarray}
\label{irrep2} \BK<(n,1)|(n,1)>&=&{n\over\ n!(n+2)}\left\{ ({\vec
u \cdot \vec u}^*)([{\vec u \times \vec R}]\cdot [{\vec u}^*\times
{\vec S}])({\vec R \cdot \vec
S})^{n-1}\phantom{n\over n}\right. \nonumber \\
&-& \left. {n-1\over n}([{\vec u \times \vec R}]\cdot [{\vec u}^*
\times  {\vec S}])^2 ({\vec R \cdot \vec S})^{n-2}\right\}
\end{eqnarray}
\begin{eqnarray}
\label{irrep3} \BK<(n-2,2)|(n-2,2)>={n-1\over2(n+1)!}([{\vec u
\times \vec R}]\cdot [{\vec u}^*\times {\vec S}])^2 ({\vec R \cdot
\vec  S})^{n-2}.
\end{eqnarray}
Note that more symmetric (in the sense of SU(3) symmetry) norm
kernels have more terms.

This brings us to the norm kernel of the $^6$He$+p$ system,
\begin{eqnarray}
\label{q11} I({\vec u},{\vec R};{\vec u}^*,{\vec S})&=& \sum_n
\lambda_n\{\BK<(n+2,0)|(n+2,0)> \nonumber \\
&&+\BK<(n,1)|(n,1)>+\BK<(n-2,2)|(n-2,2)>\}.
\end{eqnarray}
The eigenvalues $\lambda_n$ depend on $n$ but are independent of
the SU(3) indices, if $n$ is fixed. The spectrum of eigenvalues of
the norm kernel appears to be so simple because the cluster $^6$He
has no proton in its $p$-shell.

The minimal number of quanta of relative motion of the clusters is
$n_{\min}=1$. At $n\geq 1$, the irreps $(n+2,0)$ and $(n,1)$, are
realized, while the third irrep, $(n-2,2)$, appears at $n \geq 2$.
The eigenvalues $\lambda_n$ are shown in Table 1 for $n\leq11$.

\begin{table}[htb]
\label{t1}
\begin{center}
\begin{tabular}{|c|cc|}
\hline
 & $n=2k$ & $n=2k+1$  \\
 \hline
$k$ & $\lambda_n$ & $\lambda_n$ \\
\hline
0 & 0 & 1.1667 \\
1 & 0.9722 & 1.0046\\
2 & 0.9992 & 1.0001\\
3 & 0.9999 & 1.0000\\
4 & 0.9999 & 1.0000\\
5 & 1.0000 & 1.0000\\
 \hline
\end{tabular}
\end{center}
\caption{Eigenvalues $\lambda_n$ of the norm kernel for the
$^6$He+p system.}
\end{table}

To illustrate the possibility of further reductions, we project
the kernel (\ref{q11}) onto the states with $L=0$. Such states
belong to the SU(3) irreps with even $n=2k$ (k=$1,2,\dots$). The
normalized wave functions $\Psi_{(2k+2,\,0)L=0}$ and
$\Psi_{(2k-2,\,2)L=0}$ take the form
\begin{eqnarray*}
\Psi_{(2k+2,\,0)\,L=0}({\vec u},{\vec R})
=\sqrt{{1\over2(2k)!(2k+3)}} \left\{{\vec u}^2{\vec
R}^{2k}-{2k\over 2k+1}[{\vec u \times \vec R}]^2{\vec
R}^{2k-2}\right\},
\end{eqnarray*}
\begin{eqnarray*}
\Psi_{(2k-2,\,2)\,L=0}({\vec u},{\vec
R})=\sqrt{{k\over2(2k+1)!(2k+1)}}~
 [{\vec u \times \vec R}]^2{\vec R}^{2k-2}.
\end{eqnarray*}
The momentum-projected norm kernel is then
\begin{eqnarray}
I({\vec u},{\vec R};{\vec u}^*,{\vec S})&=& \sum_{k=1}^\infty
\lambda_{2k}\{\Psi_{(2k+2,\,0)L=0}({\vec u},{\vec R}) \,
\Psi_{(2k+2,\,0)L=0}({\vec u}^*,{\vec S}) \nonumber \\
&&+ \, \Psi_{(2k-2,\,2)L=0}({\vec u},{\vec R}) \,
\Psi_{(2k-2,\,2)L=0}({\vec u}^*,{\vec S})+...\},
\end{eqnarray}
where dots stand for terms with $L\ne0$.

Working with the SU(3) basis, one meets a problem of formulation
of asymptotic conditions for the wave function expansion
coefficients. Therefore, an introduction of functions of the
"physical" basis $\Phi^{(l_{\vec u},\, l_{\vec R},\, L)}_k({\vec
u},{\vec R})$, where $l_{\vec u}$ and $l_{\vec R}$ are partial
angular momenta of the $^6$He cluster and the relative motion of
the clusters, respectively, appears to be useful. The
transformation between the two bases is unitary. If $L=0$, the
"physical" functions take the form
\begin{eqnarray}
\Phi^{(0,\,0,\,0)}_k({\vec u},{\vec R})=\sqrt{{1\over
6(2k+1)!}}{\vec u}^2{\vec R}^{2k}
\end{eqnarray}
\begin{eqnarray}
\Phi^{(2,\,2,\,0)}_k({\vec u},{\vec R})=\sqrt{{2k\over
3(2k+1)!(2k+3)}} \left\{{3\over2}({\vec u\cdot \vec R})^2{\vec
R}^{2k-2}-{1\over2}{\vec u}^2{\vec R}^{2k}\right\},
\end{eqnarray}
and the transformation is
\begin{eqnarray}
\Psi_{(2k+2,0)L=0}&=&\sqrt{{2k+3\over3(2k+1)}}\Phi^{(0,0,0)}_k+
\sqrt{{4k\over3(2k+1)}}\Phi^{(2,2,0)}_k \nonumber \\
\Psi_{(2k-2,2)L=0}&=&\sqrt{{4k\over3(2k+1)}}\Phi^{(0,0,0)}_k-
\sqrt{{2k+3\over3(2k+1)}}\Phi^{(2,2,0)}_k.
\end{eqnarray}
The fact that the eigenvalues for the irreps $(2k+2,0)$ and
$(2k-2,2)$ coincide makes it possible to keep the diagonal form of
the norm kernel for this nuclear system in the "physical" basis,
\begin{eqnarray}
I({\vec u},{\vec R};{\vec u}^*,{\vec S})&=& \sum_{k=1}^\infty
\lambda_{2k}\{\Phi^{(0,\,0,\,0)}_k({\vec u},{\vec R})
\, \Phi^{(0,\,0,\,0)}_k({\vec u}^*,{\vec S}) \nonumber \\
&& \, + \,\Phi^{(2,\,2,\,0)}_k({\vec u},{\vec R})\,
\Phi^{(2,\,2,\,0)}_k({\vec u}^*,{\vec S})+...\}.
\end{eqnarray}

\subsection{System $^6$He+n}

This system is somewhat more complicated, because the dynamics of
the neutron cluster is influenced by the presence of a neutron in
the $p$-shell of $^6$He. The norm kernel now reads
\begin{eqnarray}
\label{ov6-n} I({\vec u},{\vec R};{\vec u}^*,{\vec S})&=& \sum_n
\left\{ \Lambda_{(n+2,\,0)}{1\over2\cdot n!}({\vec u \cdot \vec
u}^*)^2({\vec R \cdot \vec S})^n \right. \\
&+& \,\,\, \left. \Lambda_{(n,1)} {n(n+1)\over (n+2)!}({\vec
u\cdot \vec u}^*)([{\vec u \times \vec R}]\cdot [{\vec u}^*\times
{\vec S}]) ({\vec R \cdot \vec S})^{n-1}\right\},\nonumber
\end{eqnarray}
where the subscripts in the coefficients
\begin{eqnarray}
\Lambda_{(n+2,\,0)} =1+(-1)^n~{7n-1\over
6^n},~~~\Lambda_{(n,1)}=(-1)^{n+1}~{7(n+2)\over 2\cdot 6^n}
\end{eqnarray}
indicate the most symmetric SU(3) irrep realized in the
corresponding term of Eq.(\ref{ov6-n}). As in the previous case,
the SU(3) projection is needed, leading to the following result,
\begin{eqnarray}
\label{SU3-proj} I({\vec u},{\vec R};{\vec u}^*,{\vec S})&=&
\sum_n \{\lambda_{(n+2,0)}\BK<(n+2,0)|(n+2,0)> \\
&+&\lambda_{(n,1)}\BK<(n,1)|(n,1)>+\lambda_{(n-2,2)}\BK<(n-2,2)|(n-2,2)>\},
\nonumber
\end{eqnarray}
where
\begin{eqnarray}
\lambda_{(n+2,0)}&=&\Lambda_{(n+2,0)}, \nonumber \\
\lambda_{(n,1)}&=&\Lambda_{(n+2,0)}+\Lambda_{(n,1)}, \\
\lambda_{(n-2,2)}&=&\Lambda_{(n+2,0)}+{2(n+1)\over
n+2}\Lambda_{(n,1)}\nonumber ,
\end{eqnarray}
and the explicit form of $\BK<(\lambda,\mu)|(\lambda,\mu)>$ was
obtained above (Eqs.(\ref{irrep1}--\ref{irrep3})).

At $n=1$, only functions belonging to the space of the $(1,1)$
irrep are allowed. At $n\geq2$, all three irreps are present in
the norm kernel (see Table 2).

\begin{table}[htb]
\label{t2}
\begin{center}
\begin{tabular}{|c|ccc|ccc|}
\hline
 & \multicolumn{3}{c}{$n=2k$} & \multicolumn{2}{r}{$n=2k+1$} & \\
 \hline
$k$ & $(n+2,0)$ & $(n,1)$& $(n-2,2)$ & $(n+2,0)$ & $(n,1)$& $(n-2,2)$ \\
\hline
0 & 0 & 0 & 0  & 0 & 1.75 & 0 \\
1 & 1.3611 & 0.9722 & 0.7778  & 0.9074 & 0.9884 & 1.0370 \\
2 & 1.0208 & 1.0046 & 0.9938 & 0.9956 & 0.9988 & 1.0010 \\
3 & 1.0009 & 1.0003 & 0.9998 & 0.9998 & 0.9999 & 1.0000\\
4 & 1.0000 & 1.0000 & 0.9999 & 0.9999 & 0.9999 & 1.0000 \\
5 & 1.0000 & 1.0000 & 1.0000 & 1.0000 & 1.0000 & 1.0000 \\
 \hline
\end{tabular}
 \caption{Eigenvalues $\lambda_{(\lambda,\mu)}$ of the norm
 kernel for the system $^6$He+n.}
\end{center}
\end{table}

Again, a "physical" basis is needed to define the asymptotic form
of the wave function expansion coefficients. It is particularly
important in the calculations in the continuum, where it is
expressed in terms of the scattering $S$-matrix elements. The
eigenfunctions are the same as for $^6$He$+p$, but the unitary
transformation to the "physical" basis breaks the diagonal form of
the norm kernel\footnote{The nature of this breaking is that,
unlike the functions of the SU(3) basis, those of the "physical"
basis are not eigenfunctions of the antisymmetrization operator
and, therefore, are not invariant in respect to a permutation of
the nucleons. The permutation mixes the "physical" basis functions
with the same value of $k$. However, as $k$ increases, the degree
of mixing decreases exponentially, and at $k>5$ the norm kernel
becomes practically diagonal in the "physical" basis as well.},
due to the difference between the eigenvalues for different SU(3)
irreps. One notices, however, that this difference vanishes at
large numbers of the oscillator quanta $n=2k$, at which, indeed,
the asymptotic conditions are defined for the expansion
coefficients in the "physical" basis and related to the scattering
$S$-matrix elements. Therefore, these conditions may be converted
into the SU(3) basis. The latter has an important advantage: the
norm kernel is diagonal in the SU(3) representation. As for the
matrix elements of the Hamiltonian and, in particular, of the
kinetic energy operator, they are not more complicated when
written in the SU(3) basis.

\subsection{System $^6$He+$^4$He}

Gradually increasing the complexity of the systems studied, we
replace the neutron cluster with the $\alpha$-particle. As the
latter is free of $p$-shell nucleons, the norm kernel does not
require additional generating vectors. Consequently, the form of
both "physical" and SU(3) basis functions remains intact. The
increase in the number of nucleons will have an effect on the form
of the coefficients $\Lambda_{(\lambda,\mu)}$, and hence, of the
eigenvalues of the norm kernel.
\begin{eqnarray}
\label{const64} I({\vec u},{\vec R};{\vec u}^*,{\vec S})&=& \sum_n
\left\{ \Lambda_{(n+2,\,0)}{1\over2\cdot n!}({\vec u \cdot \vec
u}^*)^2({\vec R \cdot \vec S})^n  \right. \nonumber \\ &+&
\Lambda_{(n,1)} {n(n+1)\over (n+2)!}({\vec u \cdot \vec
u}^*)([{\vec u \times \vec R}]\cdot[{\vec u}^*\times {\vec S}])
({\vec R\cdot \vec
S})^{n-1} \nonumber \\
& +& \left. \Lambda_{(n-2,\,2)} {n-1\over 2(n+1)!}([{\vec u \times
\vec R}]\cdot[{\vec u}^*\times {\vec S}])^2 ({\vec R \cdot \vec
S})^{n-2}\right\}.
\end{eqnarray}
The explicit form of the coefficients $\Lambda_{(\lambda,\mu)}$ is
shown in Appendix B. The SU(3) projection reduces the norm kernel
to the form of Eq.(\ref{SU3-proj}), but now with
\begin{eqnarray}
\lambda_{(n+2,0)}&=&\Lambda_{(n+2,0)}, \nonumber \\
\lambda_{(n,1)}&=&\Lambda_{(n+2,0)}+\Lambda_{(n,1)}, \\
\lambda_{(n-2,2)}&=&\Lambda_{(n+2,0)}+{2(n+1)\over
n+2}\Lambda_{(n,1)}+\Lambda_{(n-2,2)}\nonumber.
\end{eqnarray}
The difference, evidently, is in the appearance of the coefficient
$\Lambda_{(n-2,\,0)}$ which was, in fact, equal to zero in the
previous case.

The values of $\lambda_{(n+2,\,2)}$ are all zeros at $n<6$. The
eigenvalues $\lambda_{(n,1)}$ equal to zero if $n<5$ and, finally,
$\lambda_{(n-2,\,2)}$ vanish for $n<4$. In other words, the states
belonging to the SU(3) irrep $(n-2,2)$ become allowed if $n\geq
4$, the minimal number of quanta for the $(\lambda,\mu)=(n,1)$
states is 5, and finally, the states with $(\lambda,\mu)=(n+2,0)$
are allowed only if $n\geq 6$ (Table 3).

\begin{table}[htb]
\label{t3}
\begin{center}
\begin{tabular}{|c|ccc|ccc|}
\hline
 & \multicolumn{3}{c}{$n=2k$} & \multicolumn{2}{r}{$n=2k+1$} & \\
 \hline
$k$ & $(n+2,0)$ & $(n,1)$& $(n-2,2)$ & $(n+2,0)$ & $(n,1)$& $(n-2,2)$ \\
\hline
0 & 0 & 0 & 0  & 0 & 0 & 0 \\
1 & 0 & 0 & 0  & 0 & 0 & 0 \\
2 & 0 & 0 & 1.2056 & 0 & 1.0549 & 0.4521 \\
3 & 0.9419 & 0.2721 & 1.1587 & 0.1831 & 1.2192 & 0.7650\\
4 & 1.2922 & 0.5698 & 1.0834 & 0.4045 & 1.1795 & 0.9011 \\
5 & 1.3264 & 0.7645 & 1.0408 & 0.5983 & 1.1160 & 0.9581 \\
6 & 1.2566 & 0.8760 & 1.0194 & 0.7448 & 1.0676 & 0.9821 \\
7 & 1.1743 & 0.9363 & 1.0090 & 0.8454 & 1.0371 & 0.9923 \\
8 & 1.1090 & 0.9678 & 1.0046 & 0.9097 & 1.0196 & 0.9967 \\
9 & 1.0645 & 0.9840 & 1.0019 & 0.9489 & 1.0101 & 0.9985 \\
10 & 1.0367 & 0.9921 & 1.0009 & 0.9718 & 1.0051 & 0.9994 \\
 \hline
\end{tabular}
\end{center}
 \caption{Eigenfunctions $\lambda_{(\lambda,\mu)}$ of the norm kernel of $^6$He$+\alpha$.}
\end{table}

\subsection{System $^8$He+$^4$He}

The next system requires three complex parameter for its norm
kernel, ${\vec u},~{\vec v}$ and ${\vec R}$. Indeed, four neutrons
in the $p$-shell of $^8$He cannot be described by the same
orbital. However, these four $p$-shell neutrons can be treated as
two neutron $p$-shell holes, which is why vectors ${\vec u},~{\vec
v}$ are entering the norm kernel only in the form of the cross
product, $[{\vec u}\times {\vec v}]\equiv{\vec w}$. Therefore, all
following expressions will contain the vectors ${\vec w}$ and
${\vec R}$ (and conjugated) only.

We again start from the norm kernel,
\begin{eqnarray}
\label{const84} I({\vec w},{\vec R};{\vec w}^*,{\vec S})&=&\sum_n
\left\{ \Lambda_{(n,\,2)}{1\over2\cdot n!}({\vec w \cdot \vec
w}^*)^2({\vec R \cdot \vec S})^n \right.
\nonumber \\
 &+& \Lambda_{(n-1,\,1)}{n\over n!(n+3)}({\vec w \cdot \vec w}^*)({\vec
w \cdot \vec R})({\vec w}^* \cdot {\vec S}) ({\vec R \cdot \vec
S})^{n-1} \nonumber \\
 &+& \left. \Lambda_{(n-2,\,0)}{(n-1)n\over 2(n+2)!}({\vec
w \cdot \vec R})^2({\vec w}^*\cdot {\vec S})^2 ({\vec R \cdot \vec
S})^{n-2}\right\}.
\end{eqnarray}
The coefficients $\Lambda_{(\lambda,\,\mu)}$ are shown in Appendix
B.

The SU(3) projection yields
\begin{eqnarray}
I({\vec w},{\vec R};{\vec w}^*,{\vec S})&=& \sum_n
\{\lambda_{(n,\,2)}\BK<(n,2)|(n,2)> + \lambda_{(n-1,\,1)}\BK<(n-1,1)|(n-1,1)> \nonumber \\
&+&\,\, \lambda_{(n-2,\,0)}\BK<(n-2,0)|(n-2,0)>\},
\end{eqnarray}
\begin{eqnarray}
\lambda_{(n,\,2)}&=&\Lambda_{(n,\,2)}, \nonumber \\
\lambda_{(n-1,\,1)}&=&\Lambda_{(n,\,2)}+\Lambda_{(n-1,\,1)}, \\
\lambda_{(n-2,\,0)}&=&\Lambda_{(n,\,2)}+{2(n+2)\over
n+3}\Lambda_{(n-1,\,1)} + \Lambda_{(n-2,\,0)} \nonumber .
\end{eqnarray}
The minimal number of quanta allowed by the Pauli exclusion
principle for the irreps $(n,2)$, $(n-1,1)$, and $(n-2,0)$ equals
6, 5, and 4, respectively (Table 4).

\begin{table}[htb]
\label{t4}
\begin{center}
\begin{tabular}{|c|ccc|ccc|}
\hline
 & \multicolumn{3}{c}{$n=2k$} & \multicolumn{2}{r}{$n=2k+1$} & \\
 \hline
$k$ & $(n,2)$ & $(n-1,1)$& $(n-2,0)$ & $(n,2)$ & $(n-1,1)$& $(n-2,0)$ \\
\hline
0 & 0 & 0 & 0  & 0 & 0 & 0 \\
1 & 0 & 0 & 0  & 0 & 0 & 0 \\
2 & 0&  0&  1.1865&  0&  0.7119& 0.6229\\
3 & 0.5006&  0.5006&  1.0197&  0.4380&  0.8551& 0.8829\\
4 & 0.7665&  0.7799&  0.9916&  0.7309&  0.9225& 0.9623\\
5 & 0.8905&  0.9064&  0.9926&  0.8797&  0.9606& 0.9873\\
6 & 0.9486&  0.9607&  0.9961&  0.9480&  0.9811& 0.9955\\
7 & 0.9762&  0.9835&  0.9982&  0.9780&  0.9912& 0.9984\\
8 & 0.9891&  0.9931&  0.9992&  0.9907&  0.9961& 0.9994\\
9 & 0.9951&  0.9971&  0.9997&  0.9961&  0.9982& 0.9998\\
10& 0.9979&  0.9988&  0.9999&  0.9984&  0.9992& 0.9999\\
 \hline
\end{tabular}
\end{center}
 \caption{Eigenvalues $\lambda_{(\lambda,\mu)}$ for the norm kernel of $^8$He$+\alpha$.}
\end{table}

The states with $L=0$ belong to the space of the $(2k,2)$ irrep
(with $k_{\min}=3$) and the $(2k-2,0)$ irrep ($k_{\min}=2$). The
wave functions of these states are
\begin{eqnarray}
\Psi_{(2k,\,2),L=0}({\vec w},{\vec
R})=\sqrt{{(k+1)^2\over(2k+3)!}} \left\{{\vec w}^2{\vec
R}^{2k}-{k\over k+1}({\vec w \cdot \vec R})^2{\vec
R}^{2k-2}\right\},
\end{eqnarray}
\begin{eqnarray}
\Psi_{(2k-2,\,0),L=0}({\vec w},{\vec R})=\sqrt{{k\over(2k+2)!}}
({\vec w \cdot \vec R})^2{\vec R}^{2k-2}.
\end{eqnarray}

The relevant part of the norm kernel reads,
\begin{eqnarray}
I({\vec w},{\vec R};{\vec w}^*,{\vec S}) &=& \sum_{k=2}^\infty
\{\lambda_{(2k,\,2)}\Psi_{(2k,\,2),L=0}({\vec w},{\vec R})
\Psi_{(2k,\,2),L=0}({\vec w}^*,{\vec S})  \\
&+& \,\, \lambda_{(2k-2,\, 0)}\Psi_{(2k-2,\, 0),L=0}({\vec
w},{\vec R}) \Psi_{(2k-2,\, 0),L=0}({\vec w}^*,{\vec S})\} + \dots
\nonumber
\end{eqnarray}

The "physical" basis states $\Phi^{(l_{\vec w},\, l_{\vec
R},L)}_k({\vec w},{\vec R})$ are labelled with the partial momenta
of the $^8$He cluster ($l_{\vec w}$) and of the relative motion of
the clusters ($l_{\vec R}$). Again, the transformation to this
basis is unitary. If $L=0$,
\begin{eqnarray}
\Phi^{(0,\,0,\,0)}_k({\vec w},{\vec R})=
{1\over\sqrt{6(2k+1)!}}{\vec w}^2{\vec R}^{2k},
\end{eqnarray}
\begin{eqnarray}
\Phi^{(2,\,2,\,0)}_k({\vec w},{\vec R})= \sqrt{{4k(k+1)\over
3(2k+3)!}}\left\{{3\over2}({\vec w \cdot \vec R})^2 {\vec
R}^{2k-2}-{1\over2}{\vec w}^2{\vec R}^{2k}\right\},
\end{eqnarray}
with
\begin{eqnarray}
\Psi_{(2k,\,2),L=0}&=&\sqrt{{2k+3\over3(k+1)}}\Phi^{(0,\,0,\,0)}_k-
\sqrt{{k\over3(k+1)}}\Phi^{(2,\,2,\,0)}_k,  \nonumber \\
\Psi_{(2k-2,\,0),L=0}&=&\sqrt{{k\over3(k+1)}}\Phi^{(0,\,0,\,0)}_k+
\sqrt{{2k+3\over3(k+1)}}\Phi^{(2,\,2,\,0)}_k .
\end{eqnarray}
The asymptotic conditions for the wave function expansion
coefficients are well known in the "physical" basis, and can be
written in the SU(3) basis using the latter relations between the
basis functions.

\section{Hamiltonian Kernel}

Besides the norm kernel, kernels of the operators entering the
Hamiltonian are needed to solve the equations of the algebraic
version of RGM. Whereas the norm kernel is the overlap integral of
the generating functions of the harmonic-oscillator basis with the
unity operator, the kernel of a physical operator is the overlap
integral of the same generating functions with this operator. The
kernel needs to be projected to the states with definite values of
oscillator quanta, indices of the SU(3) symmetry, and angular
momentum. Below we discuss the kernels of the operators of kinetic
and potential energy.

\subsection{Nucleon-Nucleon Interaction Kernel}

Consider for simplicity a central nucleon-nucleon potential having
a Gaussian form. A number of known effective potentials fall into
this category. In this case, all necessary integrations may be
done analytically, so that it remains to write down the
interaction kernel $U({\vec u}, {\vec v}, {\vec R} ;{\vec u}^*,
{\vec v}^*, {\vec S})$ in the form
\begin{eqnarray}
U({\vec u}, {\vec v}, {\vec R} ;{\vec u}^*, {\vec v}^*, {\vec S})=
\sum_{n,\tilde{n}=0}^{\infty} \, \sum_{\nu,\tilde{\nu}} \, \ME<n
\nu|\hat{U}|\tilde{n} \tilde{\nu}> \Psi_{n \nu}({\vec u}, {\vec
v}, {\vec R}) \, \tilde{\Psi}_{\tilde{n} \tilde{\nu}}({\vec u}^*,
{\vec v}^*, {\vec S}).
\end{eqnarray}
Here, $\nu$ stands for the set of all quantum numbers but the
total number of oscillator quanta $n$, i.e. the SU(3) irrep
indices $(\lambda, \mu)$, angular momentum $L$, its projection
$M$, and, where necessary, additional quantum numbers. The matrix
elements of a central interaction are diagonal over the angular
momentum and independent of its projection. The procedure of the
projection is simplified by the fact that the orthonormalized
Pauli-allowed functions $\Psi_{n \nu}$ are already found in the
Fock--Bargmann space. The interaction kernel is a bilinear in
these states form, with the matrix elements as its coefficients,
which are found by integration in the Fock--Bargmann space:
\begin{eqnarray}
\label{intu} &\ME<n \nu|\hat{U}|\tilde{n} \tilde{\nu}> = \int
d\mu_b\int d\bar{\mu}_b \,\,\, U({\vec u}, {\vec v}, {\vec R} ;
\bar{\vec{u}}^*, \bar{\vec{v}}^*, \bar{\vec{S}})\, \Psi_{n
\nu}(\bar{\vec{u}}, \bar{\vec{v}}, \bar{\vec{R}})  \,
\tilde{\Psi}_{\tilde{n} \tilde{\nu}}({\vec u}^*, {\vec v}^*, {\vec
S}) .& \nonumber \\&&
\end{eqnarray}
Note that the integration in (\ref{intu}) may be done
analytically.

Taking $^6$He$+p$ system as an example, we show a general form of
the interaction kernel $U({\vec u},{\vec R};{\vec u}^*,{\vec S})$
expanded over the $L=0$ states
\begin{eqnarray}
\lefteqn{U({\vec u},{\vec R};{\vec u}^*,{\vec S})=} && \nonumber
\\
\sum_{k,\,\tilde{k}=0}^\infty & \ME<(2k+2,0), L=0
|\hat{U}|(2\tilde{k}+2,0), L=0>
\Psi_{(2k+2,0)L=0}\tilde{\Psi}_{(2\tilde{k}+2,0)L=0}&\nonumber
\\
+&\ME<(2k+2,0), L=0|\hat{U}|(2\tilde{k}-2,2),
L=0>\Psi_{(2k+2,0)L=0} \tilde{\Psi}_{(2\tilde{k}-2,2)L=0}&
\nonumber
\\
+&\ME<(2k-2,2), L=0|\hat{U}|(2\tilde{k}+2,0),
L=0>\Psi_{(2k-2,2)L=0}\tilde{\Psi}_{(2\tilde{k}+2,0)L=0}&
\nonumber
\\
+&\ME<(2k-2,2), L=0|\hat{U}|(2\tilde{k}-2,2),
L=0>\Psi_{(2k-2,2)L=0}\tilde{\Psi}_{(2\tilde{k}-2,2)L=0}
&\nonumber
\\
+& \dots&
\end{eqnarray}
The dots here stand for the states with $L>0$.

\subsection{Kinetic Energy Kernel}

The kernel $T({\vec u}, {\vec v}, {\vec R} ;{\vec u}^*, {\vec
v}^*, {\vec S})$ of the kinetic energy of the relative motion of
clusters, as well as the related matrix elements between the basis
functions, may be obtained in the way described above. However, we
shall make use of the fact the Fock--Bargmann map $\hat{T}_{\vec
R}$ of the kinetic energy operator, defined by the relation
\begin{eqnarray}
T({\vec u}, {\vec v}, {\vec R} ;{\vec u}^*, {\vec v}^*, {\vec S})=
\hat{T}_{\vec R} \, I({\vec u}, {\vec v}, {\vec R} ;{\vec u}^*,
{\vec v}^*, {\vec S}),
\end{eqnarray}
is quite simple:
\begin{eqnarray}
\label{kin1}
 \hat{T}_{\vec R} =-{\hbar^2\over4mr_0^2}\left({\vec
R}^2-2({\vec R \cdot \vec \nabla_{\vec R} })-3+{\vec \nabla}_{\vec
R}^2\right).
\end{eqnarray}
Here $m$ is the mass of the nucleon, $r_0$ is the oscillator
length. We shall use the convention $\hbar=m=r_0=1$ from now on.

It follows from Eq.(\ref{kin1}) that the kinetic energy matrix in
the harmonic-oscillator representation is three-diagonal. The
matrix elements may be found by acting with the operator
$\hat{T}_{\vec R}$ directly on the basis states with the following
projection of the result.

To set an example, we calculate the matrix elements of the kinetic
energy operator between the $L=0$ states for the system $^6$He$+p$
(they will be valid for $^6$He+$n$ and $^8$He$+^4$He as well). We
act on the states $\Psi_{(2k+2,\,0),L=0}$ and
$\Psi_{(2k-2,\,2),L=0}$ with the first term of the operator
$\hat{T}_{\vec R}$ (\ref{kin1}) which adds two to the number of
quanta $n=2k$ of a basis function.
\begin{eqnarray}
{\vec
R}^2\Psi_{(2k+2,0),L=0}&=&\sqrt{{(2k+5)(2k+2)(2k+1)\over2k+3}}\Psi_{(2k+4,\,0),L=0}
\nonumber \\ &-& 2\sqrt{{2\over(2k+3)(2k+1)}}\Psi_{(2k,\,2),L=0},
\\
{\vec
R}^2\Psi_{(2k-2,\,2),L=0}&=&\sqrt{{2k(2k+3)^2\over2k+1}}\Psi_{(2k,\,2),L=0}.
\end{eqnarray}
Hence,
\begin{eqnarray}
\ME<(2k+4,0),L=0|\hat{T}|(2k+2,0),L=0>=-{1\over4}\sqrt{{(2k+5)(2k+2)(2k+1)\over2k+3}},
\end{eqnarray}
\begin{eqnarray}
\ME<(2k,2),L=0|\hat{T}|(2k+2,0),L=0>={1\over2}\sqrt{{2\over(2k+3)(2k+1)}},
\end{eqnarray}
\begin{eqnarray}
\ME<(2k,2),L=0|\hat{T}|(2k-2,2),L=0>=-{1\over4}\sqrt{{2k(2k+3)^2\over2k+1}}.
\end{eqnarray}
As for the diagonal matrix elements, they are generated by the
term
$${1\over2}\left(({\vec
R \cdot \vec \nabla_{\vec R}})+{3\over2}\right)$$ and depend on
the number of oscillator quanta only,
\begin{eqnarray}
\ME<(2k-2,2),L=0|\hat{T}|(2k-2,2),L=0>=&& \nonumber \\
\ME<(2k+2,0),L=0|\hat{T}|(2k+2,0),L=0> &=&{1\over2}(2k+3).
\end{eqnarray}

\subsection{Asymptotic Equations of Free Motion}

Let us expand the wave function $\Psi_{L=0}$ of a state with zero
angular momentum over the basis states $\Psi_{(2k+2,\,0)L=0}$ and
$\Psi_{(2k-2,\,2)L=0}$,
\begin{eqnarray}
\Psi_{L=0}=\sum_{k=1}^\infty C^1_k\,\Psi_{(2k+2,\,0)L=0}+
\sum_{k=1}^\infty C^2_k\,\Psi_{(2k-2,\,2)L=0},
\end{eqnarray}
where the coefficients $C^{1}_k\equiv C^{(2k+2,\,0)}_k$ and
$C^{2}_k\equiv C^{(2k-2,\,2)}_k$ satisfy the set of algebraic
linear homogeneous equations
\begin{eqnarray}
\label{u8} \sum_{\tilde{k},\tilde{\nu}}\ME<k,\nu|{\hat
H}-\lambda^\nu_{k}\,\delta_{\tilde{k},k}\,\delta_{\tilde{\nu},\nu}E|\tilde{k},\tilde{\nu}>
C^{\tilde{\nu}}_{\tilde{k}} = 0,~~{\hat H}={\hat T}+{\hat U}.
\end{eqnarray}

At $k\gg1$ the contribution of the interaction $\hat{U}$ may be
neglected, and the set of equations, reduced to
\begin{eqnarray}
\label{bb1}
\lefteqn{\ME<(2k+2,0)|\hat{T}|(2k,0)>C^1_{k-1}}&& \nonumber \\
+&\{\ME<(2k+2,0)|\hat{T}|(2k+2,0)>-\lambda_{(2k+2,0)}\, E\} C^1_k &\nonumber \\
+&\ME<(2k+2,0)|\hat{T}|(2k+4,0)>C^1_{k+1} &\nonumber \\
+& \ME<(2k+2,0)|\hat{T}|(2k,2)>C^2_{k+1} &= 0
\end{eqnarray}
\begin{eqnarray}
\label{bb2} \lefteqn{\ME<(2k-2,2)|\hat{T}|(2k,0)>C^1_{k-1}} && \nonumber \\
+& \ME<(2k-2,2)|\hat{T}|(2k-4,2)>C^2_{k-1} & \nonumber \\
+&\{\ME<(2k-2,2)|\hat{T}|(2k-2,2)>-\lambda_{(2k-2,2)}\, E\}C^2_k
& \nonumber \\
+&  \ME<(2k-2,2)|\hat{T}|(2k,2)>C^2_{k+1}
 &= 0,
\end{eqnarray}
defines the asymptotic behavior of the coefficients $C^{1,2}_k$ at
$k \gg 1$. Having expanded in Eqs.(\ref{bb1}--\ref{bb2}) the
matrix elements of ${\hat T}$ in a series over ${1/k}$ and
retaining the leading terms only, we arrive to a set of equations
which is a finite-difference representation of the Bessel
differential equations. Eq.(\ref{bb1}) is thus reduced to the
Bessel equation with the index $17/4$,
\begin{eqnarray}
\label{bes1} \left({d^2\over dy^2}+{1\over y}{d\over
dy}-{17\over4}{1\over y^2}+2E\right) C^1(y)+{2\sqrt{2}\over
y^2}C^2(y)=0;~~~y=\sqrt{2E}\,\sqrt{4k+3} ,
\end{eqnarray}
and Eq.(\ref{bb2}) is transformed into the Bessel equation with
the index $9/4$,
\begin{eqnarray}
\label{bes2} \left({d^2\over dy^2}+{1\over y}{d\over
dy}-{9\over4}{1\over y^2}+2E\right) C^2(y)+{2\sqrt{2}\over
y^2}C^1(y)=0.
\end{eqnarray}
With the matrix of the unitary transformation between the SU(3)
basis functions $\Psi_{(2k+2,\,0)L=0}$, $\Psi_{(2k-2,\,2)L=0}$,
and the "physical" basis functions
$\Phi^{(0,\,0,\,0)}_k,~\Phi^{(2,\,2,\,0)}_k$ known, we can express
the expansion coefficients $C^{1,2}_k$ in terms of
$C^{(0,\,0,\,0)}_k$ and $C^{(2,\,2,\,0)}_k$
\begin{eqnarray*}
C^1(y)=\sqrt{1\over3}C^{(0,\,0,\,0)}(y)+
\sqrt{2\over3}C^{(2,\,2,\,0)}(y),
\end{eqnarray*}
\begin{eqnarray}
\label{unit} C^2(y)=\sqrt{2\over3}C^{(2,\,2,\,0)}(y)-
\sqrt{1\over3}C^{(0,\,0,\,0)}(y).
\end{eqnarray}
It is easy to check that this orthogonal transformation splits the
set of Eqs.(\ref{bes1}--\ref{bes2}), and we arrive to two
uncoupled Bessel equations with a general solution in the form of
the Bessel and Neumann functions $J_{l+1/2}(y)$ and
$N_{l+1/2}(y)$,
\begin{eqnarray}
C^{(l_1,\,l_2,\,l)}(y)=\cos\delta_lJ_{l+1/2}(y)-\sin\delta_l
N_{l+1/2}(y),~~l=0,~2.
\end{eqnarray}
Thus we have shown that the equations for the expansion
coefficients in the SU(3) basis remain coupled even in the
asymptotic region, whereas the set of corresponding equations in
the "physical" basis is uncoupled at a large number of excitation
quanta. The asymptotic expressions for the coefficients
$C^{1,2}(y)$ easily follow those for $C^{(l_1,\,l_2,\,l)}(y)$ and
the relations (\ref{unit}).

The matrix elements of the kinetic energy operator between the
"physical" basis functions with $L=0$ have a remarkably simple
form,
\begin{eqnarray}
\ME<l_1,l_2,l,2k+2|\hat{T}|l_1,l_2,l,2k>=-{1\over4}\sqrt{(2k-l+2)(2k+l+3)},
\end{eqnarray}
\begin{eqnarray}
\ME<l_1,l_2,l,2k-2|\hat{T}|l_1,l_2,l,2k>=-{1\over4}\sqrt{(2k-l)(2k+l+1)},
\end{eqnarray}
\begin{eqnarray}
\ME<l_1,l_2,l,2k|\hat{T}|l_1,l_2,l,2k>={1\over2}\left(2k+{3\over2}\right)
.
\end{eqnarray}
Note that in the case of the system $^6$He$+p$ the identity of the
eigenvalues for all SU(3) irreps allows to employ the "physical"
basis even at small values of $k$, whereas for the other three
systems considered in this paper the "physical" basis is useful
only in defining the asymptotic behavior of the expansion
coefficients in the SU(3) basis.

\section{Conclusion}

The norm kernel of the generator-coordinate method is shown to be
a symmetric kernel of an integral equation with eigenfunctions
defined in the coordinate-momentum phase space (the Fock--Bargmann
space) and forming a complete set of orthonormalized (with the
Bargmann measure) states satisfying the Pauli exclusion principle.
The eigenvalues of the kernel of the integral equation limit to a
finite value indicating that the kernel is of special kind. The
main conclusion is that, in the Fock--Bargmann representation, the
kernel of the integral equation is always representable in the
form of a sum of degenerate kernels classified with the use of
SU(3) symmetry indices. In the absence of an SU(3) degeneracy, the
eigenspectrum of the norm kernel directly follows its form. If
there is an SU(3) degeneracy, it is found as a solution of the
integral equation which is reduced to a set of homogeneous
algebraic equations, with the rank equal to the degree of the
SU(3) degeneracy. In this way, following the requirements of the
permutational symmetry, the basis states are consistently
classified with the use of additional quantum numbers.

In order to set the asymptotic boundary conditions for the
expansion coefficients of a wave function in the SU(3) basis, a
basis with a different set of quantum numbers (the "physical"
basis) is required. The transformation between the two is defined
through a unitary matrix, and the method of its construction is
shown.

An interpretation of the norm kernel as a density matrix in the
Fock--Barg\-mann space considerably reduces the task of solving
the scattering problem for open-shell nuclei and yields a simple
result for the closeness relation for the Pauli-allowed basis
functions.

These statements were exemplified by several binary nuclear
systems with a cluster having an open $p$-shell and an
$s$-cluster. If both of the clusters have valence $p$-nucleons,
the SU(3) degeneracy occurs.  An interesting example is $^{12}$Be
with its exited states being able to decay into the
$^{6}$He$+^{6}$He and $^{8}$He$+^{4}$He channels\cite{freer}. This
system have to be studied in a coupled-channel method\cite{desc}.
Such a study in the microscopic analytical approach described here
is under its way, and its results will be published in a separate
paper.

\bigskip

\appendix
\subsection*{Appendix A: Integration in the Fock--Bargmann Space}

All necessary integrations in the Fock--Bargmann space are
performed analytically in Cartesian coordinates with the use of a
parameter differentiation. It is particularly simple to show for
functions and kernels depending on a single complex vector ${\vec
R}$ (and the conjugated vector ${\vec S}$). The Cartesian
components of these vectors are sums of their real and imaginary
parts,
$$
R_\kappa = {1\over\sqrt2} (\xi_\kappa + i \eta_\kappa), \,\,\,
S_\kappa = {1\over\sqrt2} (\xi_\kappa - i \eta_\kappa).
$$
Then,
$$
I(\alpha) \equiv {1 \over (2 \pi)^3} \int \dots \int \exp
\{-\alpha({\vec R \cdot \vec S})\} \, d\xi_1 d\xi_2 d\xi_3 d\eta_1
d\eta_2 d\eta_3
$$
$$
=  \prod_{\kappa=1}^3 {1 \over 2 \pi}\int d\xi_\kappa \int
d\eta_\kappa \exp \{- { \alpha \over 2} ( \xi^2_\kappa +
\eta^2_\kappa) \} = {1 \over \alpha^3}.
$$
In particular, the integration of the unity over the
Fock--Bargmann space (i.e. its "volume") corresponds to $I(1)=1$.

Now the result of (\ref{e1}) is obtained as
\begin{equation}
\label{app1}
 \int {1\over(2n)!}({\vec R \cdot \vec S})^{2n} d\mu_b= (-1)^n
\left[ {\partial^n\over\partial\alpha^n}I(\alpha) \right]_{\alpha
\to 0} ={(2n+1)(2n+2)\over2}.
\end{equation}

The function $\psi_{2n,0,0}$ of Eq.(\ref{ps2n0}) is normalized by
integrating an expression which depends on three independent
parameters $\alpha_{ij}$ forming a symmetric $2\times2$ matrix.
$$
I'(\alpha_{ij}) \equiv {1 \over (2 \pi)^3} \int \dots \int \exp
\{\alpha_{11} {\vec R}^2 + \alpha_{22} {\vec S}^2 + 2 \alpha_{12}
({\vec R \cdot \vec S}) \} \, d\xi_1 d\xi_2 d\xi_3 d\eta_1 d\eta_2
d\eta_3
$$
$$
=   \left({1 \over \sqrt{- \det |\alpha_{ij}|}} \right)^3,
$$
and its differentiation with respect to these parameters,
\begin{equation}
\int{\vec R}^{2n}{\vec S}^{2n}({\vec R \cdot \vec S})^m d\mu_b =
{1\over 2^m}
 \left[ {\partial^n\over\partial\alpha^n_{11}} {\partial^n\over\partial\alpha^n_{22}}
  {\partial^m\over\partial\alpha^m_{12}} I'(\alpha_{ij})
\right]_{\alpha_{ii} \to 0, \, \alpha_{12} \to -1/2} = {(2n + m+
2)! \over 2 (n+1)}.
\end{equation}
As we are interested in the $m=0$ case,
\begin{equation}
\label{app2}
 \int{\vec R}^{2n}{\vec S}^{2n} d\mu_b = (2n + 1)!
\end{equation}
In another limiting case, $n=0$, we arrive to Eq. (\ref{app1}).

\subsection*{Appendix B: Formulae for the Coefficients
$\Lambda_{(\lambda,\mu)}$}

\begin{enumerate}
\item

System $^6$He$+^4$He (see Eq.(\ref{const64}))
\begin{eqnarray}
\Lambda_{(n+2,\,0)}&=&1-{7^{n-1}\over6^n2^{n-1}}\left(5n+14-{2^{n-3}(25n^2+35n+24)\over7^{n-1}}\right)
\nonumber \\
&+&(-1)^n{2^{n-6}\over3^n}\left(25n^2-105n+64-{3^{n-2}(25n^2-70n+18)\over2^{3n-7}}\right) \\
 \Lambda_{(n,\,1)}&=& {5(n+2)2^{n-6}\over3^n}\left\{
{7^{n-1}\over8^{n-2}}\left(1-{2^{n-2}(5n+1)\over7^{n-1}}\right)
\right.
\nonumber \\
&-& \left.
(-1)^n\left(5n-13-(10n-19)\left({3\over8}\right)^{n-2}\right)\right\}
\\
\Lambda_{(n-2,\,2)}&=& {25n(n+1)\over24}\left\{{1\over6^{n-1}}+
{(-1)^n\over3}\left({2^{n-3}\over3^{n-2}}-{1\over4^{n-2}}\right)\right\}
\end{eqnarray}

\item
System $^8$He$+^4$He (see Eq.(\ref{const84}))
\begin{eqnarray}
\Lambda_{(n,\,2)}&=&1+{1\over4^{n+1}}\left(3(3n^2+9n+8)-(3n+10){5^{n-1}\over2^{n-3}}\right)
\nonumber \\
&+&
{(-1)^n\over2^{n+4}}\left(9n^2-33n+16-{9n^2-18n+2\over2^{2n-5}}\right)
\\
\Lambda_{(n-1,\,1)}&=& -{3(n+3)\over2^{n+4}}\left\{
{1\over2^{n-1}}\left(6(n+1)-{5^{n-1}\over2^{n-3}}\right) \right.
\nonumber
\\
&+& \left. (-1)^n\left(3n-7+{3(3-2n)\over4^{n-2}}\right)\right\} \\
\Lambda_{(n-2,0)}&=&{9(n+1)(n+2)\over2^{n+4}}\left\{{1\over2^{n-2}}+
(-1)^n\left(1-{1\over2^{2n-5}}\right)\right\}
\end{eqnarray}

\end{enumerate}

\end{document}